\documentclass[reprint,amsmath,aps,prc]{revtex4-2}
\usepackage{graphicx}
\usepackage{dcolumn}
\usepackage{bm}
\usepackage{hyperref}
\usepackage{amsmath}
\usepackage{amssymb}
\usepackage{amsfonts}
\usepackage{appendix}

\begin{document}
\title{Radial-flow fluctuations in the geometrical-scaling framework}
\author{Takeshi Osada}
\email{t-osada@tcu.ac.jp}
\affiliation{%
Department of Natural Sciences, Faculty of Science and Engineering, Tokyo City University,
Tamazutsumi 1-28-1, Setagaya-ku, Tokyo 158-8557, Japan.}%

\date{\today}
\begin{abstract}
We discuss radial-flow fluctuations using the 
$p_{\rm T}$-differential measure \(v_0(p_{\rm T})\), 
together with its $p_{\rm T}$-integrated counterpart \(v_0\), 
within the framework of geometrical scaling (GS), 
where the saturation momentum scale provides 
the characteristic scale for particle production. 
We show that the GS framework leads to results similar to those obtained 
from the momentum-rescaling model proposed by Jiangyong Jia. 
In the GS picture, event-by-event spectral fluctuations are 
governed by fluctuations of the saturation momentum scale; consequently, 
the single-mode ansatz introduced in Jia's model emerges naturally. 
We also show that the GS picture suggests a possible connection between 
transverse-momentum correlations and fluctuations of the emission region, 
which may be probed through Hanbury Brown and Twiss (HBT) analyses. 
Using the string percolation model, which is closely related to GS,
we estimate the multiplicity dependence of radial-flow fluctuations
and propose the scaled quantity
$A_0(N_{\Delta y}) \equiv v_0^2 N_{\Delta y}$,
with $N_{\Delta y}=(dN/dy)\Delta y$, as a diagnostic observable
for testing the role of effective flux-tube fluctuations.
\end{abstract}
\maketitle
\section{Introduction}

Recent measurements by the ATLAS Collaboration~\cite{ATLAS:2025ztg}
and the ALICE Collaboration~\cite{ALICE:2025iud} have revealed
intriguing features in the behavior of the $p_{\rm T}$-differential
radial-flow fluctuation measure $v_0(p_{\rm T})$~\cite{Mazeliauskas:2015efa}.
In particular, the ratio $v_0(p_{\rm T})/v_0$, where $v_0$ is the
$p_{\rm T}$-integrated measure, exhibits a universal shape as a
function of $p_{\rm T}$.
Although this behavior appears to be nearly universal, its physical
origin has not yet been fully understood.

Jia~\cite{Jia:2025rab} proposed a momentum-rescaling model that provides
a possible baseline for the observed shape of $v_0(p_{\rm T})$.
In this model, event-by-event spectral fluctuations are assumed to be
driven by a single collective variable, 
namely the event-by-event mean transverse momentum.
This assumption is formulated as a single-mode ansatz, 
in which the normalized spectral fluctuation is proportional to 
the fluctuation of the mean transverse momentum.
While this ansatz successfully captures important features of the data,
its physical origin remains to be clarified.

Geometrical scaling (GS) \cite{Stasto:2000er,Iancu:2002tr,Iancu:2003jg,McLerran:2014apa} 
provides a natural framework for addressing this issue.
In the GS picture, the saturation momentum scale~\cite{Kharzeev:2001gp,Gelis:2010nm} provides 
the characteristic momentum scale that governs transverse-momentum spectra.
In particular, analyses of semi-inclusive transverse-momentum spectra
in $pp$ and $p$--Pb collisions have shown that spectra over a wide range
of collision energies and multiplicities can be described in terms of a
multiplicity-dependent saturation momentum scale~\cite{Osada:2020zui,Osada:2019oor,Osada:2017oxe}.
If the GS framework is applicable to event-by-event spectral fluctuations 
in semi-inclusive events, the single-mode ansatz adopted in Jia's model can 
naturally be interpreted as arising from fluctuations of the saturation 
momentum scale, the characteristic momentum scale of the spectra.

In this paper, we show that the GS framework yields a factorized form
analogous to Jia's momentum-rescaling model.
The GS result for $v_0(p_{\rm T})$ provides a baseline associated only 
with saturation-scale fluctuations. 
Deviations from this baseline may signal additional fluctuation sources 
beyond saturation-scale fluctuations. 
We also discuss the relation between transverse-momentum correlations
and fluctuations of the transverse emission region, suggesting a
possible connection with HBT observables~\cite{Plumberg:2015mxa, Plumberg:2015aaw}.
Furthermore, the GS picture is known to be closely related to the
string percolation model~\cite{DiasdeDeus:2010ggs,Braun:2001us}.
In this model, fluctuations of the saturation momentum scale can be
interpreted in terms of fluctuations of the string density.
Using the string percolation picture, we estimate the multiplicity
dependence of radial-flow fluctuations and discuss the expected scaling
behavior of observables such as $v_0(p_{\rm T})$ and
$A_0(N_{\Delta y}) \equiv v_0^2 N_{\Delta y}$, 
with $N_{\Delta y}=(dN/dy)\Delta y$. 
This provides a phenomenological link between GS, string-density
fluctuations, and radial-flow fluctuation observables.

The paper is organized as follows.
In Sec.~II, we briefly introduce the framework of geometrical scaling
for semi-inclusive transverse-momentum spectra.
Then, we discuss how event-by-event fluctuations of produced
particle spectra are related to fluctuations of the saturation momentum
scale in semi-inclusive events.
We then show that the spectral fluctuations obtained in the GS picture
have the same single-mode structure as that assumed in Jia's model.
In Sec.~III, we derive the two-particle transverse-momentum correlation
function within the GS framework. 
Furthermore, using the known multiplicity dependence of the saturation
momentum scale, we discuss the expected $p_{\rm T}$ dependence and
multiplicity dependence of $v_0(p_{\rm T})/v_0$.
In Sec.~IV, we use the string percolation model to estimate the
multiplicity dependence of $v_0$ and related observables.
Finally, Sec.~V is devoted to a summary and concluding remarks.

\section{Transverse-Momentum Fluctuations in the Geometrical-Scaling Framework}
For each event multiplicity class, the semi-inclusive transverse momentum 
spectra of hadrons normalized by an effective cross-sectional area $S^{*}_{\rm T}$ 
can be scaled to a universal function
\begin{align}
 \frac{1}{S^{*}_{\rm T}}\frac{d N}{2\pi p_{\rm T}dp_{\rm T} dy} 
 = {\cal F}(\tau),
\end{align}
with scaling variable $\tau^{1/(2+\lambda)} = p_{\rm T}/Q_{\rm sat}$, 
where $Q_{\rm sat}$ is the multiplicity-dependent saturation momentum scale 
as a function of the effective energy $W^*$.
The function ${\cal F}(\tau)$ represents the universal scaling function, 
for which a Tsallis-type form is often used in GS analyses \cite{Osada:2020zui}:
\begin{align}
    {\cal F}(\tau) = \left[1+(q-1)\frac{\tau^{1/(2+\lambda)}}{\kappa}\right]^{\frac{-1}{q-1}},
\end{align}
where $q$ is the nonextensive parameter and $\kappa$ 
is a constant relating $Q_{\rm sat}$ to a hadronization scale, such as 
the freeze-out temperature $T_{\rm f}$. 
Denoting the event-by-event mean transverse momentum of produced charged particles by $[p_{\rm T}]$, 
its event average within a fixed multiplicity class is given by
\begin{align}
    \langle  [ p_{\rm T} ]  \rangle 
    = \frac{2\kappa}{4-3q} \langle Q_{\rm sat} \rangle.
    \label{eq:average_pt_GS}
\end{align}
Here, \( \langle Q_{\rm sat} \rangle \) denotes the event-averaged
saturation momentum scale in a given semi-inclusive event class.
Event-by-event fluctuations around this mean value are introduced below.

\subsection{Fluctuations of the saturation momentum scale and the effective interaction area}
Even within a semi-inclusive event class specified by the multiplicity, 
the event-by-event mean transverse momentum $[p_{\rm T}]$ fluctuates from event to event. 
In the GS framework, as indicated by Eq.~(\ref{eq:average_pt_GS}), 
this fluctuation can be understood as a fluctuation of the saturation momentum scale $Q_{\rm sat}$. 
Denoting the saturation momentum scale averaged over events
within a given semi-inclusive class by $\langle Q_{\rm sat}\rangle$,
we write the event-by-event saturation momentum scale as
\begin{align}
    Q_{\rm sat} = \langle Q_{\rm sat} \rangle 
    + \delta Q_{\rm sat}.
    \label{eq:Q_sat_fluctuation}
\end{align}
This saturation momentum scale is regarded 
as a function of the effective energy $W^*$ in the collision:
\begin{align}
    Q_{\rm sat}(W^*) = Q_0 
    \left(\frac{x_0 W^*}{Q_0}\right)^{\lambda/(2+\lambda)}.
\end{align}
Here, $Q_0$, $x_0$, and $\lambda$ are the constants appearing in
the saturation-momentum parametrization 
for inclusive spectra~\cite{McLerran:2014apa,Golec-Biernat:1998zce}, 
and the collision energy $W$ is replaced by the effective energy $W^*$.
Thus, within this parametrization, fluctuations of the saturation momentum scale 
reflect fluctuations of the effective energy $W^*$.
It is therefore natural to identify the mean saturation momentum scale 
entering the average spectrum of a semi-inclusive event class with 
the value evaluated at the mean effective energy 
$\langle W^* \rangle$: 
\begin{align}
 \langle Q_{\rm sat} \rangle = Q_{\rm sat}(\langle W^*\rangle).    
\end{align} 
The mean transverse momentum of charged particles produced in each event is then given by
\begin{align}
    [p_{\rm T}] &= \langle [p_{\rm T}] \rangle + \delta[p_{\rm T}] 
    = \frac{2\kappa}{4-3q} \left( \langle Q_{\rm sat}\rangle +\delta Q_{\rm sat} \right) \nonumber \\
    &= \langle [p_{\rm T}] \rangle \left(1 + 
    \frac{\delta Q_{\rm sat}}{\langle Q_{\rm sat}\rangle}\right), 
    \label{eq:[p_T]}
\end{align}
where $\langle [p_{\rm T}] \rangle$ denotes the average of the event-by-event mean transverse momentum
over events in a given semi-inclusive class. 
It corresponds to the mean transverse momentum 
$\langle p_{\rm T} \rangle$ for that semi-inclusive class. 
The corresponding transverse-momentum fluctuation is
\begin{align}
       \frac{\delta [p_{\rm T}]}{\langle [p_{\rm T}] \rangle} 
    =\frac{\delta Q_{\rm sat}}{\langle Q_{\rm sat}\rangle}
    \approx \frac{\lambda}{2+\lambda} \frac{\delta W^*}{\langle W^* \rangle}.
    \label{eq:delta_pt_delta_Qsat}
\end{align}
Furthermore, for semi-inclusive transverse momentum spectra in the
rapidity interval \(-\Delta y/2 < y < +\Delta y/2\), 
\begin{align}
    N(p_{\rm T}) \equiv \frac{d N}{
    dp_{\rm T} dy}, 
\end{align}
the multiplicity in this interval,
\(N_{\Delta y}\equiv (dN/dy)\Delta y\), is fixed. Therefore,
\begin{align}
    N_{\Delta y} =\int dp_{\rm T} N(p_{\rm T}) ~
    =
    \frac{2\pi \kappa^2~\Delta y}{(2-q)(3-2q)}
    S^{*}_{\rm T} Q_{\rm sat}^2.
    \label{eq:dn_ch_dy_GS}
\end{align}
This relation implies a correlation between \(Q_{\rm sat}\) and
\(S^*_{\rm T}\).
If fluctuations of the other spectral parameters, such as $\kappa$ and $q$, can be neglected,
\begin{align}
    \frac{\delta S^{*}_{\rm T}}{\langle S^{*}_{\rm T}\rangle} 
    = -2 \frac{\delta Q_{\rm sat}}{\langle Q_{\rm sat}\rangle} . 
    \label{eq:deltaS_dettaQ} 
\end{align}
Thus, Eq.~(\ref{eq:deltaS_dettaQ}) shows that fluctuations of the 
saturation momentum scale are directly related to fluctuations of the effective area $S^{*}_{\rm T}$.

\subsection{Fluctuations of the transverse-momentum spectrum in the GS framework}
The average transverse-momentum spectrum over a semi-inclusive event class, 
namely an ensemble of events satisfying a fixed multiplicity condition 
$dN/dy$ around midrapidity, may be approximated by using event-averaged 
saturation momentum $\langle Q_{\rm sat}\rangle$ and the event-averaged 
effective area $\langle S^{*}_{\rm T}\rangle$ as follows:
\begin{align}
\langle N(p_{\rm T}) \rangle
&\approx N(p_{\rm T}, \langle S^{*}_{\rm T}\rangle, \langle Q_{\rm sat}\rangle) \nonumber \\
&= 2\pi p_{\rm T}
   \langle S^{*}_{\rm T}\rangle
   \left[ 1+(q-1)\frac{p_{\rm T}}{\kappa \langle Q_{\rm sat}\rangle} \right]^{\frac{-1}{q-1}}.
   \label{eq:average_spectrum_GS}
\end{align}
Note that, in general,
\begin{align}
\langle N(p_{\rm T}) \rangle
&\neq N(p_{\rm T}, \langle S^{*}_{\rm T}\rangle, \langle Q_{\rm sat}\rangle). 
\end{align}
This distinction should be noted when event-by-event fluctuations are included. 
The approximation becomes valid when the fluctuations around the mean spectrum are small 
and second-order effects in the spectral fluctuations can be neglected. 
Such second-order effects enter naturally in the two-particle correlation functions 
discussed below. 

The spectrum in an event where the saturation momentum and the effective 
area fluctuate around their respective event-averaged values is written as
\begin{align}
N(p_{\rm T})  &= N(p_{\rm T}, \langle S^{*}_{\rm T}\rangle 
+ \delta S^{*}_{\rm T}, \langle Q_{\rm sat}\rangle + \delta Q_{\rm sat}) . 
\end{align}
For each event, the mean transverse momentum of 
produced charged particles is given by Eq.~(\ref{eq:[p_T]}).
Assuming that the event-by-event deviation of the transverse-momentum
spectrum from the event-averaged spectrum is small, we expand the spectral
fluctuation to first order:
\begin{align}
& N(p_{\rm T})
 \approx N(p_{\rm T}, \langle S^{*}_{\rm T}\rangle, \langle Q_{\rm sat}\rangle)  \nonumber \\
&+ \frac{\partial N(p_{\rm T}, S^{*}_{\rm T}, Q_{\rm sat})}{\partial S^{*}_{\rm T}}
\bigg|_{\langle S^{*}_{\rm T}\rangle, \langle Q_{\rm sat}\rangle}  \times 
\delta S^{*}_{\rm T} \nonumber \\
& + \frac{\partial N(p_{\rm T}, S^{*}_{\rm T}, Q_{\rm sat})}{\partial Q_{\rm sat}}
\bigg|_{\langle S^{*}_{\rm T}\rangle, \langle Q_{\rm sat}\rangle} \times
\delta Q_{\rm sat} \nonumber \\
&\approx  \langle N(p_{\rm T}) \rangle
\left[1+\frac{\frac{p_{\rm T}}{\kappa \langle Q_{\rm sat}\rangle}}
{ 1+(q-1)\frac{p_{\rm T}}{\kappa \langle Q_{\rm sat}\rangle} }
\frac{\delta Q_{\rm sat}}{\langle Q_{\rm sat}\rangle} 
+\frac{\delta S^{*}_{\rm T}}{\langle S^{*}_{\rm T}\rangle} \right]. 
\end{align}
(For compactness, we use the notation
$\left.\cdots\right|_{\langle S^{*}_{\rm T}\rangle,\langle Q_{\rm sat}\rangle}$
to indicate evaluation at
$(S^{*}_{\rm T},Q_{\rm sat})
=(\langle S^{*}_{\rm T}\rangle,\langle Q_{\rm sat}\rangle)$.)
Using $\delta N(p_{\rm T})\equiv N(p_{\rm T})-\langle N(p_{\rm T})\rangle$ and Eq.~(\ref{eq:deltaS_dettaQ}), one obtains
\begin{align}
    \frac{\delta N(p_{\rm T})}{\langle N(p_{\rm T}) \rangle} 
    &\approx 
    \left[
    \frac{\frac{p_{\rm T}}{\kappa \langle Q_{\rm sat}\rangle }}{ 1+(q-1)\frac{p_{\rm T}}{\kappa \langle Q_{\rm sat}\rangle } }
    -2 \right]
    \frac{\delta Q_{\rm sat}}{\langle Q_{\rm sat}\rangle} 
    \label{eq:fluctuation_GS_Qsat}, 
\end{align}
and using Eq.~(\ref{eq:[p_T]}), this can also be written as
\begin{align}
    \frac{\delta N(p_{\rm T})}{\langle N(p_{\rm T}) \rangle} 
    &\approx 
    \left[
    \frac{\frac{p_{\rm T}}{\kappa \langle Q_{\rm sat}\rangle }}
    { 1+(q-1)\frac{p_{\rm T}}{\kappa \langle Q_{\rm sat}\rangle } }
    -2 \right]
    \frac{\delta [p_{\rm T}]}{\langle [p_{\rm T}] \rangle}.
    \label{eq:fluctuation_GS}
\end{align}

\subsection{Comparison with Jia's momentum-rescaling model}
We now examine the same normalized spectral fluctuation using Jia's rescaling model.
Here the event average is again understood as an average over 
a multiplicity-fixed semi-inclusive event class. As assumed in Jia's 
rescaling model, the transverse-momentum spectrum observed 
in each event is determined by the event-by-event mean transverse 
momentum $[p_{\rm T}]$ and by particle-number conservation. 
In other words, each event spectrum is generated from the ensemble-averaged
spectrum by the rescaling
\[
p_{\rm T} \to p_{\rm T}\frac{\langle [p_{\rm T}] \rangle}{[p_{\rm T}]},
\]
together with the corresponding normalization factor.
Following the notation of Jia~\cite{Jia:2025rab}, we introduce the
event-wise normalized transverse-momentum spectrum
\footnote{
Since the multiplicity is fixed in a semi-inclusive event class,
\(N(p_{\rm T})\) and the normalized spectrum \(n(p_{\rm T})\)
differ only by an overall constant. We therefore use \(N(p_{\rm T})\)
in the definition of \(v_0(p_{\rm T})\) below, while the corresponding
single-mode relation is equivalently written for \(n(p_{\rm T})\).
}
as
\begin{align}
    n(p_{\rm T}) \equiv \frac{1}{N_{\Delta y}}N(p_{\rm T}),
    \qquad
    N_{\Delta y} \equiv \frac{dN}{dy}\Delta y ,
\end{align}
where \(\Delta y\) is the width of the rapidity window. The event average
of \(n(p_{\rm T})\) is denoted by
\begin{align}
    f(p_{\rm T}) \equiv \langle n(p_{\rm T}) \rangle
    =
    \frac{1}{N_{\Delta y}}\langle N(p_{\rm T})\rangle .
\end{align}
We assume that the spectrum in each event is generated by rescaling
the transverse-momentum axis of the average spectrum \(f(p_{\rm T})\):
\begin{align}
    n(p_{\rm T})
    &=
    \frac{\langle [p_{\rm T}] \rangle}{[p_{\rm T}]}
    f\left(
    p_{\rm T}
    \frac{\langle [p_{\rm T}] \rangle}{[p_{\rm T}]}
    \right)
    \nonumber \\
    &=
    \frac{1}{1+\delta[p_{\rm T}]/\langle [p_{\rm T}] \rangle}
    f\left(
    \frac{p_{\rm T}}
    {1+\delta[p_{\rm T}]/\langle [p_{\rm T}] \rangle}
    \right) .
\end{align}
Here the prefactor
\(\langle [p_{\rm T}] \rangle/[p_{\rm T}]
=1/(1+\delta[p_{\rm T}]/\langle [p_{\rm T}] \rangle)\)
is introduced so that the normalization condition of the spectrum is
preserved under the rescaling of the transverse-momentum axis:
\begin{align}
    \int dp_{\rm T}\, n(p_{\rm T})
    =
    \int dp_{\rm T}\, f(p_{\rm T})
    = 1 .
    \label{eq:normalization_Jia}
\end{align}
Substituting the geometrical-scaling form of the average spectrum,
Eq.~(\ref{eq:average_spectrum_GS}), into the rescaling relation
and expanding the resulting expression to first order in 
$\delta[p_{\rm T}]/\langle [p_{\rm T}] \rangle$, one obtains
\begin{align}
    n(p_{\rm T})&\approx 
    \frac{2\pi p_{\rm T} 
    \langle S^{*}_{\rm T} \rangle}{N_{\Delta y}}
    \left[ 1+(q-1)\frac{p_{\rm T}}{\kappa \langle Q_{\rm sat}\rangle} \right]^{\frac{-1}{q-1}} 
    \nonumber \\ 
    &\times \left[ 1 - 2\frac{\delta[p_{\rm T}]}
    {\langle [p_{\rm T}] \rangle} \right]
    \left[
        1+\frac{\frac{p_{\rm T}}{\kappa \langle Q_{\rm sat}\rangle}}
        { 1+(q-1)\frac{p_{\rm T}}{\kappa \langle Q_{\rm sat}\rangle} }
    \frac{\delta[p_{\rm T}]}{\langle [p_{\rm T}] \rangle} \right]. 
    \end{align}
Therefore, the relative fluctuation of the transverse-momentum spectrum 
at fixed $p_{\rm T}$, i.e., the vertical fluctuation, is obtained as
\begin{align}
    \frac{\delta n(p_{\rm T})}{\langle n(p_{\rm T}) \rangle} 
     &\approx 
    \left[
    \frac{\frac{p_{\rm T}}{\kappa \langle Q_{\rm sat}\rangle }}
    { 1+(q-1)\frac{p_{\rm T}}{\kappa \langle Q_{\rm sat}\rangle } }
    -2 \right]
    \frac{\delta[p_{\rm T}]}{\langle [p_{\rm T}] \rangle}. 
\end{align}
This is identical to Eq.~(\ref{eq:fluctuation_GS}), 
which was derived for fluctuations of semi-inclusive 
transverse momentum spectra in the GS framework.

\section{Two-particle momentum correlation and radial flow fluctuations in the GS picture}
\subsection{Two-particle transverse-momentum correlation}
The correlation function of semi-inclusive transverse momentum spectra in the GS framework can be obtained directly from the spectral fluctuations. We define the correlation function as
\begin{align}
C(p_{\rm T1},p_{\rm T2}) &\equiv 
\frac{\langle N(p_{\rm T1}) N(p_{\rm T2}) \rangle}
{\langle N(p_{\rm T1}) \rangle \langle N(p_{\rm T2}) \rangle} -1 \nonumber \\ 
&= \frac{\langle \delta N(p_{\rm T1}) \delta N(p_{\rm T2}) \rangle}
{\langle N(p_{\rm T1}) \rangle \langle N(p_{\rm T2}) \rangle}. 
\end{align}
Here, the GS spectral response function $K_{\rm GS}(p_{\rm T})$ is defined as
\begin{align}
    K_{\rm GS}(p_{\rm T}) \equiv \frac{\frac{p_{\rm T}}{\kappa \langle Q_{\rm sat}\rangle }}
{ 1+(q-1)\frac{p_{\rm T}}{\kappa \langle Q_{\rm sat}\rangle } }-2 . 
\label{eq:def_of_K_GS}
\end{align}
Using this definition together with Eq.~(\ref{eq:fluctuation_GS_Qsat}), we obtain
\begin{align}
&C(p_{\rm T1},p_{\rm T2}) = \frac{K_{\rm GS}(p_{\rm T1}) K_{\rm GS}(p_{\rm T2})}
{\langle Q_{\rm sat}\rangle^2} \langle \left(\delta Q_{\rm sat} \right)^2\rangle . 
\label{eq:correlation_GS}
\end{align}
Equation~(\ref{eq:correlation_GS}) indicates that the correlation function 
of semi-inclusive transverse momentum spectra in the GS framework 
is determined by the magnitude of the saturation momentum fluctuation, 
$\langle(\delta Q_{\rm sat})^2\rangle$, and by the 
spectral response function $K_{\rm GS}(p_{\rm T})$.

Since Eq.~(\ref{eq:deltaS_dettaQ}) holds in semi-inclusive events,   
if we define a size scale of the effective area 
by $R^*_{\rm T} \equiv \sqrt{S^{*}_{\rm T}/\pi}$, then 
the fluctuation of $R_{\rm T}^*$ is related to the fluctuation of $Q_{\rm sat}$ as
\begin{align}
    \frac{\delta R_{\rm T}^*}{\langle R_{\rm T}^* \rangle} 
    = - \frac{\delta Q_{\rm sat}}{\langle Q_{\rm sat}\rangle}. 
\end{align}
Thus, in the GS framework, the transverse-momentum correlation function is also related to fluctuations of the size scale associated with the effective area:
\begin{align}
&C(p_{\rm T1},p_{\rm T2}) = 
K_{\rm GS}(p_{\rm T1}) 
K_{\rm GS}(p_{\rm T2}) 
\frac{\langle (\delta R_{\rm T}^*)^2\rangle}{\langle R^{*}_{\rm T}\rangle^2}.
\end{align}
The quantity $\langle(\delta R_{\rm T}^*)^2\rangle/\langle {R}^{*}_{\rm T}\rangle^2$ 
is expected to be constrained, for example, through HBT analyses of the emission region. 
Although $R_{\rm T}^*$ is not identical to an HBT radius, the GS picture predicts 
a possible connection between transverse-momentum correlations and fluctuations of 
the emission region inferred from HBT measurements~\cite{Plumberg:2015mxa, Plumberg:2015aaw}:
\begin{align}
\frac{\langle (\delta R_{\rm T}^*)^2\rangle}{\langle R^{*}_{\rm T}\rangle^2}=
\frac{C(p_{\rm T1},p_{\rm T2})}{K_{\rm GS}(p_{\rm T1}) K_{\rm GS}(p_{\rm T2})}.
\end{align}

Thus, the GS picture suggests that the spectral correlation function is
connected not only to saturation-scale fluctuations but also to
fluctuations of the effective transverse area.
The corresponding size scale, $R_{\rm T}^*$, however, should not be
identified directly with the HBT radius $R_{\rm HBT}$.
Clarifying the relation between these two quantities remains an
important subject for future study.

\subsection{$p_{\rm T}$-differential and -integrated measure of radial-flow fluctuations}
The $p_{\rm T}$-differential measure of radial-flow fluctuations,
$v_0(p_{\rm T})$, quantifies the component of the yield fluctuation
at a given $p_{\rm T}$ that is correlated with the event-by-event mean
transverse momentum $[p_{\rm T}]$. 
The $p_{\rm T}$-differential measure $v_0(p_{\rm T})$ is defined as
\begin{align}
    v_0(p_{\rm T})&\equiv \rho(N(p_{\rm T}),[p_{\rm T}])
    \frac{\sqrt{\langle (\delta N(p_{\rm T}))^2 \rangle}}{\langle N(p_{\rm T}) \rangle}, 
    \nonumber \\
    &=\frac{\langle \delta N(p_{\rm T}) \delta [p_{\rm T}] \rangle}
    {\langle N(p_{\rm T}) \rangle \sqrt{\langle (\delta [p_{\rm T}])^2 \rangle}},
\end{align}
where
\begin{align}
\rho(N(p_{\rm T}),[p_{\rm T}]) \equiv 
    \frac{\langle \delta N(p_{\rm T}) \delta [p_{\rm T}] \rangle}
    {\sqrt{\langle (\delta N(p_{\rm T}))^2 \rangle \langle (\delta [p_{\rm T}])^2 \rangle}}
\end{align}
is the Pearson correlation coefficient between the spectrum fluctuation 
$\delta N(p_{\rm T})$ and the mean transverse momentum fluctuation $\delta [p_{\rm T}]$.
In the geometrical-scaling framework,
this correlation can be interpreted as the response of the spectrum to
event-by-event fluctuations of the saturation momentum scale $Q_{\rm sat}$.
This interpretation is supported by the fact that, to leading order in
$\delta Q_{\rm sat}$, one obtains
\begin{align}
    &\langle \delta N(p_{\rm T}) \delta [p_{\rm T}] \rangle 
    =\langle N(p_{\rm T}) \rangle \langle [p_{\rm T}] \rangle 
    K_{\rm GS}(p_{\rm T}) \frac{\langle (\delta Q_{\rm sat})^2 \rangle}
    {\langle {Q}_{\rm sat}\rangle^2}.
\end{align}
In this approximation, both $\delta N(p_{\rm T})$ and $\delta[p_{\rm T}]$
are driven by the same event-by-event fluctuation $\delta Q_{\rm sat}$.
Therefore, the magnitude of the Pearson correlation coefficient is unity,
\begin{align}
\left|\rho(N(p_{\rm T}),[p_{\rm T}])\right|=1,
\end{align}
while its sign is determined by $K_{\rm GS}(p_{\rm T})$. Substituting the
above covariance into the definition of $v_0(p_{\rm T})$, we obtain
\begin{align}
v_0(p_{\rm T}) =
K_{\rm GS}(p_{\rm T})
\frac{\sqrt{\langle (\delta Q_{\rm sat})^2 \rangle}}
{\langle {Q}_{\rm sat}\rangle}.
\end{align}
The relation $\left|\rho(N(p_{\rm T}),[p_{\rm T}])\right|=1$ means 
that the spectral fluctuation and the mean-transverse-momentum 
fluctuation are perfectly correlated or anticorrelated. 
In the present GS picture, 
both originate from the same physical source, namely 
the event-by-event fluctuation of the saturation momentum scale 
$Q_{\rm sat}$. 
This also implies that $v_0(p_{\rm T})$ is determined by the 
spectral response function 
$K_{\rm GS}(p_{\rm T})$ and by the magnitude of the saturation-scale fluctuation, 
$\sqrt{\langle(\delta Q_{\rm sat})^2\rangle}/\langle Q_{\rm sat} \rangle$.
On the other hand, the
$p_{\rm T}$-integrated measure $v_0$ is defined as
\begin{align}
    v_0 &\equiv 
     \int dp_{\rm T} 
    \left[ \frac{p_{\rm T}-\langle [p_{\rm T}] \rangle}{\langle [ p_{\rm T} ] \rangle} \right]
    v_0(p_{\rm T}) ~\langle n(p_{\rm T}) \rangle 
     \nonumber \\
    &=\frac{\sqrt{\langle (\delta [p_{\rm T}])^2 \rangle}}{\langle [p_{\rm T}] \rangle}
    = \frac{\sqrt{\langle (\delta Q_{\rm sat})^2 \rangle}}
    {\langle {Q}_{\rm sat}\rangle}.
\end{align}
Therefore,
\begin{align}
    \frac{v_0(p_{\rm T})}{v_0} = K_{\rm GS}(p_{\rm T}).
    \label{eq:v0_pt/v_0}
\end{align}
In the GS picture, we therefore obtain
\begin{align}
    \frac{\delta n(p_{\rm T})}{\langle n(p_{\rm T}) \rangle}
    &= K_{\rm GS}(p_{\rm T}) \frac{\delta [p_{\rm T}]}{\langle [p_{\rm T}] \rangle} \nonumber \\
    &= \frac{v_0(p_{\rm T})}{v_0} \frac{\delta [p_{\rm T}]}{\langle [p_{\rm T}] \rangle}. 
\end{align}
This is precisely the single-mode ansatz assumed by Jia in the rescaling model [Eq.~(2) of Ref.~\cite{Jia:2025rab}]. Thus, in the minimal GS picture considered here, the single-mode ansatz is not an additional assumption but follows naturally when the dominant fluctuation source is the saturation momentum scale.

\subsection{GS baseline for $v_0(p_{\rm T})/v_0$ }
We now examine the ratio of the $p_{\rm T}$-differential and
$p_{\rm T}$-integrated measures, $v_0(p_{\rm T})/v_0$, obtained in the GS
picture. 
From Eq.~(\ref{eq:v0_pt/v_0}), this ratio reduces simply to
$K_{\rm GS}(p_{\rm T})$, namely the quantity defined in
Eq.~(\ref{eq:def_of_K_GS}) as a function of $p_{\rm T}$.
For semi-inclusive events, this result represents the contribution
arising solely from fluctuations of the saturation momentum scale.
Therefore, deviations of the measured $v_0(p_{\rm T})/v_0$ from 
$K_{\rm GS}(p_{\rm T})$ may indicate the presence of fluctuation sources 
not accounted for by saturation-scale fluctuations. 
In this sense, $K_{\rm GS}(p_{\rm T})$ can serve as a baseline 
for isolating such additional contributions.

According to our previous work~\cite{Osada:2020zui}, 
the saturation momentum scale $Q_{\rm sat}$ and the effective interaction 
size scale $R_{\rm T}^* $ 
depend on the multiplicity fixed in the semi-inclusive event class, 
namely the value of $dN/dy$ at midrapidity. They can be parametrized as
\begin{subequations}
\begin{align}
    Q_{\rm sat} &= a_Q + b_Q \left(\frac{dN}{dy}\right)^{1/6},  
    \label{eq:Q_sat_1/6_power}\\
    R_{\rm T}^* &= a_R + b_R \left(\frac{dN}{dy}\right)^{1/3}.
\end{align}
\end{subequations}
The values of $a_Q$, $b_Q$, $a_R$, and $b_R$ for $pp$ and $p$--Pb collisions at several collision energies are listed in Table~\ref{tab:fit_parameters}.

\begin{table}[t]
\caption{
Parameters used to parametrize the multiplicity-dependent saturation
momentum $Q_{\rm sat}$ according to Eq.~(\ref{eq:Q_sat_1/6_power}).
In the numerical estimate, we use $q=1.145$ and $\kappa=0.1100$,
which are the values of the universal function used for the GS
analysis of semi-inclusive $\pi^\pm$ spectra in $pp$ and $p$--Pb
collisions. 
The values are reproduced in part from Table~II of
Ref.~\cite{Osada:2020zui}.
}
\label{tab:fit_parameters}
\begin{ruledtabular}
\begin{tabular}{lcc}
System and energy &
$a_Q$ & $b_Q$
\\
\hline
\multicolumn{3}{l}{$pp\to \pi^\pm+X$}
\\
$2.76~{\rm TeV}$~\cite{CMS:2012xvn}
&
$-0.019$ & $0.854$
\\
$7.00~{\rm TeV}$~\cite{CMS:2012xvn}
&
$-0.149$ & $0.954$
\\
$7.00~{\rm TeV}$~\cite{ALICE:2018pal}
&
$-0.225$ & $0.985$
\\
$13.0~{\rm TeV}$~\cite{CMS:2017eoq}
&
$-0.472$ & $1.164$
\\[1mm]
\multicolumn{3}{l}{$p{\rm -Pb}\to \pi^\pm+X$}
\\
$5.02~{\rm TeV}$~\cite{CMS:2013pdl}
&
$0.078$ & $0.899$
\\
$5.02~{\rm TeV}$~\cite{ALICE:2013wgn}
&
$0.315$ & $0.600$
\\
\end{tabular}
\end{ruledtabular}
\end{table}

\begin{figure*}[t]
  \centering
  \includegraphics[
    width=0.75\textwidth,
    height=0.35\textheight,
    keepaspectratio
  ]{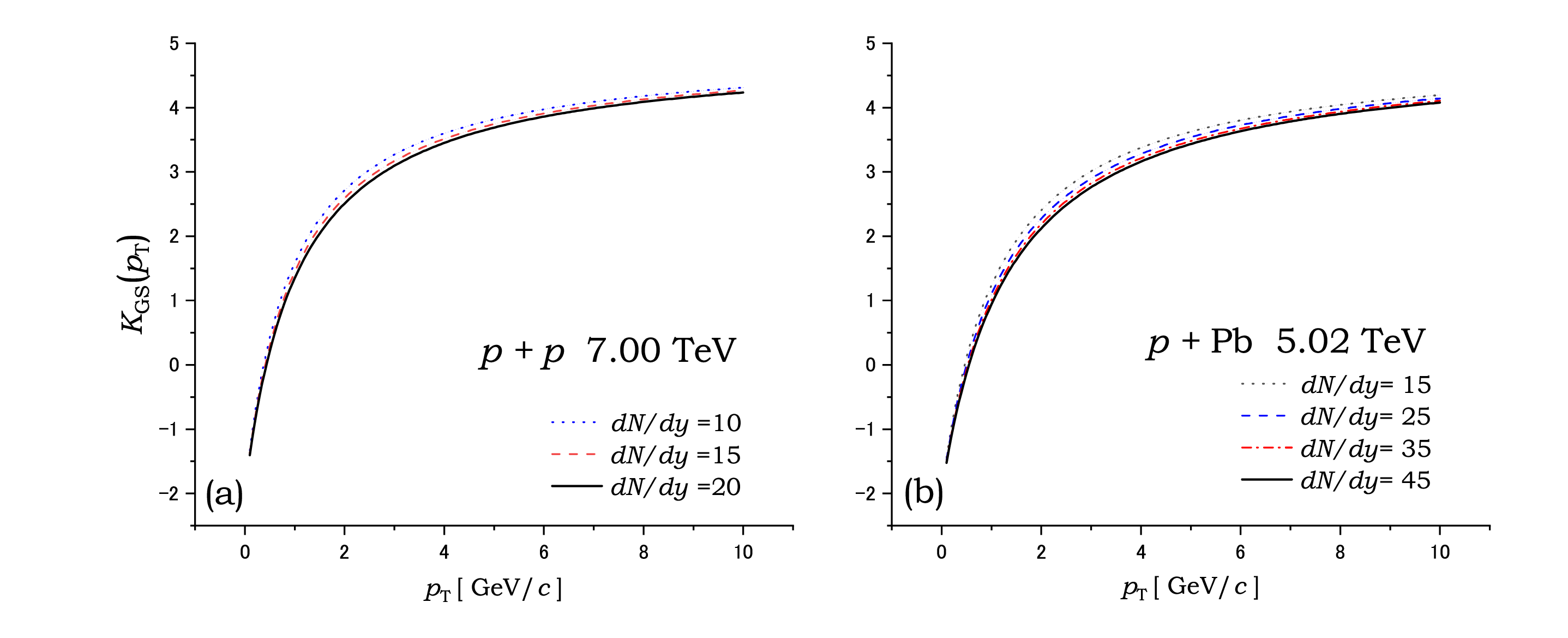}
  \caption{
    Baseline estimate of $K_{\rm GS}(p_{\rm T})$ for $pp$ and $p$--${\rm Pb}$ collisions
    obtained from the geometric-scaling parametrization.
    The Tsallis parameters $q=1.145$ and $\kappa=0.1100$ are fixed to the values used in
    Table~\ref{tab:fit_parameters}. 
    The curves represent the spectral response to fluctuations of
    $Q_{\rm sat}$ within the GS framework, with no additional 
    dynamical fluctuation sources included.
  }
\label{fig:KGS_pp_pPb_baseline}
\end{figure*}

As representative examples, we consider $pp\to \pi^{\pm}+X$ at $\sqrt{s}=7.00$~TeV~\cite{ALICE:2018pal} and $p$--Pb$\to \pi^{\pm}+X$ at $\sqrt{s}=5.02$~TeV~\cite{CMS:2013pdl}. The resulting $K_{\rm GS}$ is shown in Fig.~\ref{fig:KGS_pp_pPb_baseline}.
As shown in Fig.~\ref{fig:KGS_pp_pPb_baseline}, 
\(K_{\rm GS}(p_{\rm T})\) starts from a negative value at low \(p_{\rm T}\), 
crosses zero at around \(p_{\rm T}\simeq 0.5\)--\(0.7\) GeV/\(c\), 
and increases monotonically with increasing \(p_{\rm T}\).\footnote{
It should be noted that the zero-crossing point is not determined solely
by the spectral response itself, but can also be affected by analysis
conventions, such as the finite \(p_{\rm T}\) range used to define
\(v_0\) and the normalization of the event-by-event spectra.}
The obtained GS spectral response function exhibits a similar
\(p_{\rm T}\) dependence for \(pp\) and \(p\)--Pb collisions, suggesting
a universal character of the response function within the present GS
framework.

It should be emphasized that the present minimal GS baseline does not
produce the high-\(p_{\rm T}\) falloff observed in heavy-ion collision
data. 
Rather, it accounts for the sign change and the low-\(p_{\rm T}\) rise of
\(v_0(p_{\rm T})/v_0\) associated with the spectral response to
saturation-scale fluctuations. 
A deviation of the measured \(v_0(p_{\rm T})/v_0\) from this monotonic baseline, 
especially at high \(p_{\rm T}\), may indicate the presence of additional fluctuation
sources, such as those arising from jet production or other hard
processes, which are not included in the minimal GS picture.

\section{Relation to the string percolation model}
The string percolation model (see Refs.~\cite{DiasdeDeus:2010ggs,Braun:2001us} 
and the Appendix of Ref.~\cite{Osada:2020zui}) also predicts that the spectral 
fluctuation and the mean-transverse-momentum fluctuation are strongly correlated, 
in close analogy with the GS picture. 

The area covered by strings is given by \(S^*_{\rm T}(1-e^{-\eta})\), 
where \(S^*_{\rm T}e^{-\eta}\) represents the uncovered area. 
The average number of effective color flux tubes is then estimated by 
dividing this covered area by the area of an effective string, 
\(\sigma_1 F(\eta)\):
\begin{align}
    \langle N_{\rm eff} \rangle
    = \frac{S^*_{\rm T}(1-e^{-\eta})}{\sigma_1 F(\eta)} .
\end{align}
Here, \(\eta\) is the mean string density, defined as 
\(\eta = N_{\rm s}\sigma_1/S^*_{\rm T}\), with \(N_{\rm s}\) being the 
number of strings. 
The area of a single string is denoted by \(\sigma_1\), and the factor 
\(F(\eta)\) accounts for the reduction of the effective string area due 
to the overlap of strings in the transverse plane, which is given by
\begin{align}
    F(\eta) = \sqrt{\frac{1-e^{-\eta}}{\eta}} .
\end{align}
The quantity \(\langle N_{\rm eff}\rangle\) may be regarded as the 
counterpart of \(S^*_{\rm T}Q_{\rm sat}^2\), which represents the average 
number of color flux tubes in the GS picture. 
Equating these two quantities gives
\begin{align}
    Q_{\rm sat}^2 
    = \frac{\sqrt{\eta(1-e^{-\eta})}}{\sigma_1} 
    \propto 
    \begin{cases}
    \eta, & \eta \ll 1, \\
    \sqrt{\eta}, & \eta \gg 1 .
    \end{cases}
\end{align}
In the high-density limit $\eta\gg1$, the effective energy dependence of the mean 
string density $\eta$ may be estimated as
\begin{align}
    \eta= (\sigma_1 Q_0^{\frac{2}{2+\lambda}})^2 {W^{*}}^{\frac{2\lambda}{2+\lambda}}, 
\end{align}
and, for $\lambda=0.22$, one obtains $\eta \sim (W^{*})^{1/5}$.
The fluctuation of the saturation momentum scale can then be related to the fluctuation 
of the string density $\eta$ as
\begin{align}
    \frac{\delta Q_{\rm sat}}{\langle Q_{\rm sat} \rangle}
    = \frac{\alpha(\eta)}{4}  \frac{\delta \eta}{\eta}, 
    \label{eq:deltaQ_delta_eta}
\end{align}
where $\alpha(\eta)$ is a function of $\eta$ defined as
\begin{align}
    \alpha(\eta)
    &\equiv 1+\frac{\eta e^{-\eta}}{1-e^{-\eta}},
    \label{eq:alpha_eta}\\
    \alpha(\eta)
    &\to
    \begin{cases}
        2, & \eta \ll 1,\\
        1, & \eta \gg 1.
    \end{cases}
    \nonumber
\end{align}
As shown in Eq.~(\ref{eq:deltaQ_delta_eta}), 
the string percolation model also leads to a strong correlation 
between the spectral fluctuation and the mean-transverse-momentum fluctuation via the 
common source of the fluctuation of the saturation momentum scale $\delta Q_{\rm sat}$. 
Hence, using Eqs.~(\ref{eq:fluctuation_GS_Qsat}) and (\ref{eq:deltaQ_delta_eta}), 
the relative fluctuation of the spectrum at fixed $p_{\rm T}$ can be expressed as
\begin{align}
    \frac{\delta N(p_{\rm T})}{\langle N(p_{\rm T}) \rangle} 
    = K_{\rm GS}(p_{\rm T}) \frac{\alpha(\eta)}{4}  \frac{\delta \eta}{\eta}
    .  
\end{align}
Squaring Eq.~(\ref{eq:deltaQ_delta_eta}) and taking the event average gives
\begin{equation}
  \left\langle
  \left(
  \frac{\delta Q_{\rm sat}}{Q_{\rm sat}}
  \right)^2
  \right\rangle
  =
  \frac{\alpha^2(\eta)}{16}
  \left\langle
  \left(
  \frac{\delta\eta}{\eta}
  \right)^2
  \right\rangle .
\end{equation}
If the fluctuations of the string density are governed by the finite
number of effectively independent flux tubes, \(N_{\rm eff}\), the
relative variance is expected to scale as
\begin{equation}
  \left\langle
  \left(
  \frac{\delta\eta}{\eta}
  \right)^2
  \right\rangle
  \sim
  \frac{1}{N_{\rm eff}}, 
\end{equation}
and then we have
\begin{equation}
  \left\langle
  \left(
  \frac{\delta Q_{\rm sat}}{Q_{\rm sat}}
  \right)^2
  \right\rangle
  \sim
  \frac{\alpha^2(\eta)}{16N_{\rm eff}} .
\end{equation}
This relation provides a simple estimate of the magnitude of
saturation-scale fluctuations in terms of the number of effectively
independent flux tubes.
%
In the geometric-scaling description of semi-inclusive spectra, 
the multiplicity $N_{\Delta y}$ is given by Eq.~(\ref{eq:dn_ch_dy_GS}). 
Since the number of effective flux tubes is also estimated as
$N_{\rm eff}\sim S^*_T Q_{\rm sat}^2$, we may write phenomenologically
\begin{equation}
  N_{\rm eff}
  =
  C_{\rm eff} N_{\Delta y}, 
\end{equation}
where \(C_{\rm eff}\) is an unknown constant of order unity. This constant absorbs the difference between the number of effective flux tubes and the observed charged-particle multiplicity.
Therefore,
\begin{equation}
  \left\langle
  \left(
  \frac{\delta Q_{\rm sat}}{Q_{\rm sat}}
  \right)^2
  \right\rangle
  \sim
  \frac{\alpha^2(\eta)}{16~C_{\rm eff}  N_{\Delta y}}.
\end{equation}
In the geometric-scaling picture, 
the integrated radial-flow fluctuation \(v_0\) may then be estimated as
\begin{equation}
  v_0^2
  \equiv
  \left\langle
  \left(
  \frac{\delta [p_T]}{\langle [p_T]\rangle}
  \right)^2
  \right\rangle
  \simeq
  \left\langle
  \left(
  \frac{\delta Q_{\rm sat}}{Q_{\rm sat}}
  \right)^2
  \right\rangle .
\end{equation}
Thus, we obtain the following estimate for \(v_0^2\):
\begin{equation}
  v_0^2
  \sim
  \frac{\alpha^2(\eta)}{16C_{\rm eff}}
  \frac{1}{N_{\Delta y}}.
\end{equation}
This motivates the following scaled observable:
\begin{equation}
  A_0(N_{\Delta y}) \equiv
   v_0^2 N_{\Delta y}.
\end{equation}
In the string percolation picture, however, \(A_0\) can be suppressed
at high multiplicity through the density dependence of \(\alpha(\eta)\),
which reflects the nonlinear reduction associated with string fusion
or percolation. 
\begin{equation}
  A_0(N_{\Delta y})
  \sim
  \frac{\alpha^2(\eta)}{16~C_{\rm eff}} =
  \begin{cases}
    \displaystyle
    \frac{1}{4C_{\rm eff}},
    & N_{\Delta y}\sim 1 ~(\eta \ll 1), \\[10pt]
    \displaystyle
    \frac{1}{16C_{\rm eff}},
    & N_{\Delta y}\gg 1 ~(\eta \gg 1).
  \end{cases}
\end{equation}
Thus, the observable  $A_0(N_{\Delta y})$
is a useful diagnostic quantity. If the system is described by a simple 
independent-source picture with a fixed relation between \(N_{\rm eff}\) 
and multiplicity, this quantity should be approximately constant. 
On the other hand, if the system evolves from the low-density regime,
$Q_{\rm sat}^2 \propto \eta$,
to the high-density percolated regime, $Q_{\rm sat}^2 \propto \sqrt{\eta}$,
then one expects a suppression of $v_0^2 N_{\Delta y}$ 
toward high multiplicity. In the simplest estimate, 
the high-density value is smaller than the low-density 
value by a factor of four:
\begin{equation}
\frac{
    \left[ A_0(N_{\Delta y}) \right]_{N_{\Delta y}\gg 1}
  }
  { \left[ A_0(N_{\Delta y}) \right]_{N_{\Delta y}\sim {\cal O}(1)} 
  }
   \sim
  \frac{1}{4}.
\end{equation}
This behavior would indicate the nonlinear reduction of density fluctuations 
associated with string fusion or percolation.

\section{Summary and Concluding Remarks}
We have discussed radial-flow fluctuations in the framework of
geometrical scaling. In semi-inclusive event classes with fixed
multiplicity, fluctuations of the saturation momentum scale
\(Q_{\rm sat}\) are correlated with fluctuations of the effective
interaction area \(S^*_{\rm T}\). As a consequence, the spectral
fluctuation of the transverse-momentum spectrum at fixed \(p_{\rm T}\)
can be written in terms of a single fluctuation mode,
\(\delta Q_{\rm sat}/\langle Q_{\rm sat}\rangle\), which is directly
related to the mean-transverse-momentum fluctuation
\(\delta[p_{\rm T}]/\langle[p_{\rm T}]\rangle\).
This structure is precisely the single-mode ansatz assumed by Jia
in the momentum-rescaling model~\cite{Jia:2025rab}. Thus, the GS
picture provides a natural explanation for the single-mode structure
of radial-flow fluctuations. The spectral response function
\(K_{\rm GS}(p_{\rm T})\) then serves as a baseline for isolating
additional contributions to radial-flow fluctuations beyond those
arising from saturation-scale fluctuations.

We have also discussed the two-particle transverse-momentum
correlation function in the GS framework. The correlation function
factorizes into the product of \(K_{\rm GS}(p_{\rm T1})\) and
\(K_{\rm GS}(p_{\rm T2})\), multiplied by the variance of the
saturation-momentum fluctuation,
\(\langle(\delta Q_{\rm sat})^2\rangle/\langle Q_{\rm sat}\rangle^2\).
Through the fixed-multiplicity constraint, this variance is also
related to fluctuations of the effective transverse size. This result
suggests a possible connection between transverse-momentum correlations
and fluctuations of the emission region inferred from HBT measurements.

Finally, we have discussed the relation between the GS picture and
the string percolation model. In this picture, saturation-scale
fluctuations can be interpreted as fluctuations of the string density
\(\eta\). Assuming that the relative variance of the string density is
governed by the number of effective flux tubes, \(N_{\rm eff}\), we
obtained a simple estimate of the magnitude of saturation-scale
fluctuations in terms of \(N_{\rm eff}\). This estimate motivates the
scaled observable
\[
A_0(N_{\Delta y}) \equiv v_0^2 ~N_{\Delta y}, 
\qquad N_{\Delta y}=\frac{dN}{dy}\Delta y,
\]
which is expected to be approximately constant in a simple
independent-source picture, but can be suppressed at high multiplicity
in the string percolation picture due to the nonlinear reduction of
density fluctuations associated with string fusion or percolation.
The measurement of \(A_0(N_{\Delta y})\) and its multiplicity dependence would
provide a useful diagnostic for understanding the underlying physics
of radial-flow fluctuations and the possible role of string
percolation in high-energy collisions.

The present GS baseline should be regarded as the minimal spectral response
generated by saturation-scale fluctuations at fixed multiplicity.
Therefore, a comparison of this baseline with the measured
\(v_0(p_{\rm T})/v_0\) can isolate deviations that are not accounted for
by the GS response alone.  
Such deviations may provide insight into the presence of additional
fluctuation sources beyond saturation-scale fluctuations, and may contain
information about the dynamics of multiparticle production and possible
final-state effects reflected in the observed spectra and correlations.

\bibliography{osada_2604fin}
\end{document}